\documentclass[twocolumn,showpacs,amsmath,amssymb,prd]{revtex4}
\usepackage{graphicx}
\usepackage{dcolumn}
\usepackage{bm}

\def\beq{\begin{equation}}
\def\enq{\end{equation}}
\def\ba{\begin{eqnarray}}
\def\ea{\end{eqnarray}}
\def\<{<\!\!}
\def\>{\!\!>}
\def\ra{\rightarrow}
\def\eps{\epsilon}
\def\vareps{\varepsilon}

\begin{document}
\input{epsf}

\title{Enhancement of the $\bar\nu_e$ flux from astrophysical sources by 
two photon annihilation interactions}

\author{Soebur Razzaque,$^{1,2}$ Peter M\'esz\'aros$^{1,2}$ and 
Eli Waxman$^3$}

\affiliation{$^1$Department of Astronomy \& Astrophysics, 
Pennsylvania State University, University Park, PA 16802, USA}
\affiliation{$^2$Department of Physics, 
 Pennsylvania State University, University Park, PA 16802, USA}
\affiliation{$^3$Physics Faculty, Weizmann Institute of Science, 
Rehovot 76100, Israel}

\begin{abstract}
The ratio of anti-electron to total neutrino flux,
$\Phi_{\bar{\nu}_e}:\Phi_\nu$, expected from $p\gamma$ interactions in
astrophysical sources is $\le1:15$. We point out that this ratio is
enhanced by the decay of $\mu^+\mu^-$ pairs, created by the
annihilation of secondary high energy photons from the decay of the
neutral pions produced in $p\gamma$ interactions. We show that, under
certain conditions, the $\Phi_{\bar{\nu}_e}:\Phi_\nu$ ratio may be
significantly enhanced in gamma-ray burst (GRB) fireballs, and that
detection at the Glashow resonance of $\bar{\nu}_e$ in kilometer scale
neutrino detectors may constrain GRB fireball model parameters, such
as the magnetic field and energy dissipation radius.
\end{abstract}


\date{\today}
\maketitle

\section{Introduction} 

High energy neutrinos are expected to be produced in astrophysical
sources mainly by $p\gamma$ interactions, leading to the production
and subsequent decay of charged pions: $\pi^+ \ra e^{+} \nu_e {\bar
\nu}_{\mu} \nu_{\mu}$ (see, e.g., \cite{revs} for recent reviews). 
Neutrino oscillations lead in this case to an observed ratio of
$\bar\nu_e$ flux to the total $\nu$ flux of $\simeq 1:15$
\cite{Learned:1994wg} (or lower, in case muons suffer significant
electromagnetic energy loss prior to decay \cite{kashti05}). For
neutrinos produced in inelastic $pp$ ($pn$) nuclear collisions, where
both $\pi^+$'s and $\pi^-$'s are produced, the ratio is $\simeq 1:6$,
and it was suggested that measurements of the $\nu_e$ to $\bar\nu_e$
flux ratio at the $W$-resonance may allow one to probe the physics of
the sources by discriminating between the two primary modes of pion
production, $p\gamma$ and $pp$ collisions
\cite{Anchordoqui:2004eb}. This test for discriminating between the
two mechanisms is complicated by the fact that the ratio of
$\Phi_{\bar{\nu}_e}$ to $\Phi_\nu$ produced in $p\gamma$ interactions
can be enhanced to a value similar to that due to inelastic nuclear
collisions in sources where the optical depth to $p\gamma$
interactions is large (e.g. \cite{kashti05}). In this case, neutrons
produced in $p\gamma\rightarrow n\pi^+$ interactions are likely to
interact with photons and produce $\pi^{-}$ before escaping the
source, leading to production of roughly equal numbers of $\pi^{+}$'s
and $\pi^{-}$'s.

In this paper, we point out that the $\Phi_{\bar{\nu}_e}:\Phi_\nu$
ratio from $p\gamma$ interactions may be enhanced above $1:15$ also in
sources with small $p\gamma$ optical depth. Neutral pions, which are
created at roughly the same rate as charged pions in $p\gamma$
interactions, decay to produce high energy $\gamma$-rays. These
$\gamma$-rays typically carry $\sim10\%$ of the initial proton energy,
and may therefore interact with the low energy photons (with which the
protons interact to produce pions) to produce $\mu^+\mu^-$ pairs. The
decay of muons yields (after vacuum oscillations)
$\Phi_{\bar{\nu}_e}:\Phi_\nu\simeq1:5$, thus enhancing the $\bar\nu_e$
fraction.

We discuss below a specific example, the widely considered fireball
model of GRBs. In this model, the observed $\gamma$-rays are produced
by synchrotron radiation of shock accelerated electrons in the
magnetic field which is assumed to be a fraction of the total energy
(see \cite{zhang04} e.g. for reviews). The protons are expected to
co-accelerate with electrons to ultra-high energy \cite{wax95}, and
produce high energy neutrinos by $p\gamma$ interactions \cite{wb97}.
We calculate below the additional neutrino flux, due to the decay of
muons produced by secondary photon annihilation, for a typical long
duration GRB, and show that the enhanced $\bar\nu_e$ flux may be
detectable at the Glashow resonance (${\bar \nu}_e e \ra W^- \ra {\rm
anything}$ \cite{glashow:60}) in kilometer scale neutrino detectors
such as IceCube \cite{ahrens04}.

The enhancement of $\Phi_{\bar{\nu}_e}:\Phi_\nu$ due to $\gamma\gamma$
interactions in $p\gamma$ sources makes the discrimination between
$p\gamma$ and $pp$ neutrino sources more difficult.  On the other
hand, it may provide a new handle on the physics of the source.  We
show below that for GRBs the enhancement of ${\bar \nu}_e$ flux
depends on model parameters which are poorly constrained by
observations, namely the magnetic field strength and the energy
dissipation radius. Detection of ${\bar \nu}_e$'s at the Glashow
resonance, in conjunction with $\gamma$-ray detection, may therefore
constrain these parameters.

\section{Fireball model and photon spectrum} 

The minimum observed GRB fireball radius $r$ may be estimated by
requiring that it is optically thin to Thomson scatterings:
$\tau'_{\rm Th} = \sigma_{\rm Th} n' r' \lesssim 1$ (denoting the
comoving and local lab. frame variables with and without a prime
respectively). Here $n'$ is the density of scatterers in the fireball,
$r'=r/\Gamma$ is the size of the interaction region and $\Gamma$ is
the bulk Lorentz factor. The radius at which $\tau'_{\rm Th} \approx
1$ is the photospheric radius $r_{\rm ph}$. For a kinetic luminosity
$L_{\rm k}$ of the fireball, mostly carried by the baryons, the number
density of the baryons, and of the leptons which are coupled to the
baryons, is $n'_{\rm b} \approx L_{\rm k}/(4\pi r^2 \Gamma^2
m_pc^3)$. The observed isotropic equivalent $\gamma$-ray luminosity of
a long duration GRB is $L_{52}=L_{\gamma}/10^{52}{\rm
erg/s}\sim1$. Assuming $L_{\gamma}= \vareps_{e} L_{\rm k}$ with
$\vareps_{e} \sim 0.05 \vareps_{e,-1.3}$ (a parametrization which is
motivated below), the photospheric radius is
\ba 
r_{\rm ph} = \frac{\sigma_{\rm Th} L_{\gamma}/\vareps_{e}} {4\pi
\Gamma^3 m_p c^3} \approx 7.4\times 10^{12}
~\frac{L_{52}}{\vareps_{e,-1.3} \Gamma_{2.5}^{3}}~{\rm cm},
\label{photo-baryon} 
\ea
for $\Gamma_{2.5} =\Gamma/316\sim 1$. The radius at which the bulk
kinetic energy dissipation occurs, e.g. by internal shocks, is in
general $r \gtrsim r_{\rm ph}$.

The $\gamma$-ray spectrum of a GRB fireball at a dissipation radius
$r=10^{14} r_{14}$ cm peaks at a typical energy
\ba 
\eps_{\gamma,\rm pk} &=& \hslash c \Gamma^2
(3\gamma_{e,\rm min}^{'2} q B')/(2m_e c^2) \nonumber \\ &\sim & 500
\left( \vareps_{e,-1.3}^3 \vareps_{B,-1} L_{52}
\Gamma_{2.5}^2 /r_{14}^2 \right)^{1/2} {\rm keV},
\label{synpeak-energy} 
\ea
due to synchrotron radiation by electrons with a Lorentz factor
$\gamma'_{e,\rm min} \approx \vareps_e (m_p/m_e)$.  Here,
$\gamma'_{e,\rm min}$ is at the lower end of a $\propto
1/\gamma_e^{'p}$ distribution of electron Lorentz factor, with
$p\gtrsim 2$, created by Fermi acceleration in the shock. The magnetic
field is assumed to be $B^{'2}/8\pi \approx \vareps_B L_{\rm k}/(4\pi
r^2 \Gamma^2 c)$, where $\vareps_B \sim 0.1 \vareps_{B,-1}$ is the
equipartion value, currently unconstrained in the GRB prompt
phase. Note that $\eps_{\gamma,\rm pk} \propto 1/r$ will be larger
than the above value for $r \approx r_{\rm ph}$, with other parameters
fixed.

For a GRB at a luminosity distance $d_L$ the observed $\gamma$-ray
spectrum is generally approximated with a broken power-law Band fit
\cite{band93},
\ba
\frac{dN_{\gamma}}{d\eps_{\gamma}} \approx \frac{L_{\gamma}} 
{4\pi d_{L}^2 \eps_{\gamma,\rm pk}^2} \begin{cases}
(\eps_{\gamma}/\eps_{\gamma,\rm pk})^{-1} ~;~ 
\eps_{\gamma} < \eps_{\gamma,\rm pk} \cr
(\eps_{\gamma}/\eps_{\gamma,\rm pk})^{-2} ~;~ 
\eps_{\gamma} > \eps_{\gamma,\rm pk}. \end{cases}
\label{Band-fit}
\ea
The spectrum deviates from this at low energy, becoming
$dN_{\gamma}/d\eps_{\gamma} \propto \eps_{\gamma}^{3/2}$ for
$\eps_{\gamma} \lesssim \eps_{\gamma,\rm sa}$, the energy below which
synchrotron self-absorption becomes dominant.  Theoretical modeling
indicates a value \cite{razzaque04b}
\ba
\eps_{\gamma,\rm sa} &\approx &
2.4 \left( \Gamma^2 \gamma'_{e,\rm min} n'_{e} r [q\hslash c]^4
B^{'2}/[m_e^3 c^6] \right)^{1/3} \nonumber \\ &\sim & 8
\left(\vareps_{B,-1} L_{52}^2/ [\vareps_{e,-1.3} 
\Gamma_{2.5}^2 r_{14}^3] \right)^{1/3} {\rm keV},
\label{sa-freq-baryon}
\ea
for $p=2$. The differential number density of photons is
\ba 
dN'_{\gamma}/d\eps'_{\gamma} \approx L_{\gamma}/(4\pi r^2 c
\eps_{\gamma,\rm pk}^2) ~~~~~~~~~~~~~~~~~~~~~~~~~~~~~~~~
\nonumber \\ ~~~~~~~~\times
\begin{cases} \left( \eps'_{\gamma,\rm sa}
/\eps'_{\gamma,\rm pk} \right)^{-1} \left(
\eps'_{\gamma} /\eps'_{\gamma,\rm sa} \right)^{3/2} ;
\eps'_{\gamma} < \eps'_{\gamma,\rm sa} \cr \left(
\eps'_{\gamma} /\eps'_{\gamma,\rm pk} \right)^{-1} ;
\eps'_{\gamma,\rm pk} > \eps'_{\gamma} > \eps'_{\gamma,\rm sa} \cr
\left( \eps'_{\gamma} /\eps'_{\gamma,\rm pk} \right)^{-2} ;
\eps'_{\gamma} > \eps'_{\gamma,\rm pk}. \end{cases}
\label{gamma-co-dens-all} 
\ea 

Electron synchrotron radiation produces a power law $\gamma$-ray
spectrum at energies above $\eps_{\gamma,\rm pk}$ [see
Eq. (\ref{synpeak-energy})] which depends on the maximum Lorentz
factor. Other mechanisms can contribute to an extension of the
$\gamma$-ray spectrum in Eq. (\ref{Band-fit}) to high energies.  High
energy electrons can inverse Compton scatter synchrotron photons up to
an energy similar to the maximum shock accelerated electron energy
(which we derive shortly) in the Klein-Nishina limit in one
mechanism. Here we consider ultra-high energy $\gamma$-rays from
$\pi^{0}$ decays which are produced by $p\gamma$ interactions of shock
accelerated protons with synchrotron photons as $p\gamma \ra \Delta^+
\ra p\pi^0 \ra p \gamma\gamma$.

The maximum proton energy is calculated by equating its acceleration
time $t'_{\rm acc} \approx \eps'_{p}/(qcB')$ to the shorter of the
dynamic time $t'_{\rm dyn} \approx r/(2\Gamma c)$ and the synchrotron
cooling time $t'_{\rm syn} \approx 6\pi m_{p}^4 c^3/(\sigma_{\rm Th}
m_e^2 \eps'_{p} B^{'2})$ as
\ba
\eps_{p,\rm max} &=& \left( 6\pi m_{p}^4 c^4 q \Gamma^2
/[\sigma_{\rm Th} m_e^2 B'] \right)^{1/2} \nonumber \\ &\approx &
3.3\times 10^{11} \left( \vareps_{e,-1.3} \Gamma_{2.5}^6 r_{14}^2
/[\vareps_{B,-1} L_{52}] \right)^{1/4} {\rm GeV} \nonumber \\
\eps_{p,\rm max} &=& \frac{q B' r}{2} \approx 
5.5\times 10^{11} \left( \frac{\vareps_{B,-1}
L_{52}}{\vareps_{e,-1}\Gamma_{2.5}^2}
\right)^{1/2} {\rm GeV},
\label{max-p-energy}
\ea
respectively for $t'_{\rm acc} = t'_{\rm syn}$ and $t'_{\rm acc} =
t'_{\rm dyn}$. For electrons, $t'_{\rm dyn} \gg t'_{\rm syn}$
typically and the maximum electron energy, using
Eq. (\ref{max-p-energy}) for electrons, is $\eps_{e,\rm max} \approx
10^6 \vareps_{e,-1.3}^{1/4} \vareps_{B,-1}^{-1/4} L_{52}^{-1/4}
\Gamma_{2.5}^{3/2} r_{14}^{1/2}$ GeV.

At a given incident proton energy $\eps_{p}$, the threshold photon
energy leading to a $\Delta^+$ resonance interaction is
\ba
\eps_{\gamma,\Delta^+} \simeq 
\frac{0.3 \Gamma^2}{(\eps_{p}/{\rm GeV})} ~{\rm GeV}. 
\label{delta-resonance}
\ea
The optical depth for this interaction may be calculated using a
delta-function approximation with a cross-section $\sigma_{p\gamma}
\approx 10^{-28}$ cm$^2$ as
\ba
\tau'_{p\gamma} (\eps'_{p}) = \sigma_{p\gamma} \frac{r}{\Gamma} 
\left( \frac{dN'_{\gamma,\Delta^+}}{d\eps'_{\gamma,\Delta^+}} \right)
d\eps'_{\gamma,\Delta^+}
\label{pgamma-opacity}
\ea
using Eq. (\ref{gamma-co-dens-all}). Note that the target photon
spectrum, within parenthesis, is now evaluated at the $\Delta^+$
resonance energy [see Eq. (\ref{delta-resonance})] for an incident
proton energy $\eps'_p$. In particular, Eq. (\ref{delta-resonance})
may be used to replace $\eps'_{\gamma,\Delta^+}$ by $\eps'_p$ in
Eq. (\ref{gamma-co-dens-all}). As a result, the optical depth in
Eq. (\ref{pgamma-opacity}) is expressed as a function of
$\eps'_p$. The spectral shape of the optical depth is then $\propto
(\eps'_p)^{q-1}$, where $q$ is the spectral index,
$dN'_{\gamma}/d\eps'_{\gamma} \propto (\eps'_{\gamma})^{-q}$, in
Eq. (\ref{gamma-co-dens-all}). Also the order is reversed,
i.e. $q=2,1,-3/2$ in $\tau'_{p\gamma}\propto (\eps'_p)^{q-1}$ instead
of $q=-3/2,1,2$ in $dN'_{\gamma}/d\eps'_{\gamma} \propto
(\eps'_{\gamma})^{-q}$, according to the condition in
Eq. (\ref{delta-resonance}).

Protons lose $\approx 20\%$ of their energy by $p\gamma$ interactions
to $\pi^{0}$ and $\eps_{\gamma} \approx 0.1 \eps_{p}$ for each
secondary photon. With an equal probability to produce $\pi^{0}$ and
$\pi^{+}$ in each $p\gamma$ interaction, the resulting $\pi^0$ decay
photon flux is
\ba
\frac{dN_{\gamma}}{d\eps_{\gamma}} = 
{\rm min}[1,\tau'_{p\gamma}]
\frac{0.2}{4} \frac{(\xi_p/\vareps_e) L_{\gamma}} 
{4\pi d_L^2 \eps_{\gamma}^2}.
\label{pi0-gamma-flux}
\ea
Here $\xi_p$ is the proton fraction undergoing shock acceleration. For
$\xi_p =1$ and $\vareps_e =0.05$ we have $\tau'_{p\gamma} \approx 1$
which leads to the observed flux level in Eq. (\ref{Band-fit}).
Secondary pions from $\Delta^+$ decay, and subsequent decay photons
and neutrinos follow the $dN_p/d\eps_p \propto \eps_{p}^{-p}$ spectral
shape of the protons for a constant optical depth. For an optical
depth of spectral shape $\propto \eps_{p}^{q-1}$, the resulting pion,
photon and neutrino spectra would be $dN/d\eps \propto
\eps^{q-1-p}$. The $\pi^0$ decay photon spectrum in 
Eq. (\ref{pi0-gamma-flux}) is then $dN_{\gamma}/d\eps_{\gamma}
\propto \eps_{\gamma}^{-2}$ between $\eps_{\gamma} =
0.03\Gamma^2/\eps_{\gamma,\rm pk} ~{\rm GeV^2}$ and
$0.03\Gamma^2/\eps_{\gamma,\rm sa} ~{\rm GeV^2}$, $\propto
\eps_{\gamma}^{-1}$ below $\eps_{\gamma} = 0.03\Gamma^2/
\eps_{\gamma,\rm pk} ~{\rm GeV^2}$ and $\propto
\eps_{\gamma}^{-9/2}$ above $\eps_{\gamma} = 0.03\Gamma^2/
\eps_{\gamma,\rm sa} ~{\rm GeV^2}$ due to self-absorption
following Eqs. (\ref{gamma-co-dens-all}) and (\ref{pgamma-opacity}).

Note that, the luminosity of shock-accelerated protons is $1/\vareps_e
= 20$ times the shock-accelerated electron luminosity. In the fast
cooling scenario, valid in the GRB internal shocks, the electrons
synchrotron radiate all their energy into observed $\gamma$-rays.
Thus $L_{p} \approx L_{\gamma}/\vareps_e$. In a single $p\gamma$
interaction the secondary pion (charged or neutral) luminosity is
$L_{\pi} \approx 0.2 L_{p} \approx 4 L_{\gamma}/\vareps_{e,-1.3}$.
The neutrino luminosity of all flavors from charged pion decay,
assuming equal energy for all 4 final leptons, is $L_{\nu} \approx
(1/2)(3/4) L_{\pi} \approx 1.5 L_{\gamma}/\vareps_{e,-1.3}$. The 1/2
factor arises from the equal probability of $\pi^0$ and $\pi^+$
production.  The neutrinos carry away energy from the fireball, and
the rest of the pion decay ($e^+$ from $\pi^+$ and $\gamma\gamma$ from
$\pi^0$) energy is electromagnetic (e.m.), with a luminosity $L_{\rm
e.m.} \approx L_{\pi}-L_{\nu} \approx 2.5
L_{\gamma}/\vareps_{e,-1.3}$.  A significant fraction of the
$\pi^0$-decay $L_{\gamma}$ [Eq. (\ref{pi0-gamma-flux})] would be
converted to muon pairs and subsequently to neutrinos, as we discuss
next. A substantial (small) fraction of the rest of $L_{\rm e.m.}$
would be emitted at $r>r_{\rm ph}$ ($r\approx r_{\rm ph}$) as low
energy photons with a luminosity not significantly above the observed
$\gamma$-ray luminosity. These, however, do not affect substantially
the neutrino flux calculated from very high energy $\gamma$-rays
interacting with soft photons.

\section{Two Photon Pair Production} 

High energy $\gamma$-rays can produce lepton pairs, $l^+l^-$
($l=e,\mu$), with other photons which are above a threshold energy
$\omega_{\rm th} =m_lc^2$ in the center of mass (c.m.) frame of
interaction. For an incident (target) photon of energy
$\eps'_{\gamma,\rm i}$ ($\eps'_{\gamma,\rm t}$) in the comoving GRB
fireball frame, $\omega =(2\eps'_{\gamma,\rm i}
\eps'_{\gamma,\rm t})^{1/2}$, and the cross-section for $l^+l^-$ pair
production may be written, ignoring the logarithmic rise factor at
high energy, as $\sigma_{\gamma\gamma \ra l^+l^-} \approx \pi r_e^2
(m_l c^2/\omega)^2$, where $r_e$ is the classical electron radius. The
corresponding optical depth is
\ba
\tau_{\gamma\gamma} (\eps'_{\gamma,\rm i}) =
\frac{r}{\Gamma} \int \sigma_{\gamma\gamma} (\eps'_{\gamma,\rm i};
\eps'_{\gamma,\rm t}) \frac{dN'_{\gamma,\rm t}}
{d\eps'_{\gamma,\rm t}} d\eps'_{\gamma,\rm t}.
\label{opt-co-define}
\ea
Given the power-law dependence of the photon distribution in
Eq. (\ref{gamma-co-dens-all}), we may calculate the $l^+l^-$ pair
production opacities by integrating Eq. (\ref{opt-co-define})
piecewise as
\ba
\tau_{\gamma\gamma \ra l^+l^-} (\eps_{\gamma}) =
r_e^2 m_e^2 c^3 L_{\gamma}/
(8r\eps_{\gamma,\rm pk}^2 \eps_{\gamma}) ~~~~~~~~~~~~~~~~~~~
\nonumber \\ ~~~ \times
\begin{cases}
\frac{1}{2} \left[ \left( \frac{\eps_{\gamma,\rm pk}} 
{\eps_{l,\rm th}} \right)^2 - \left(\frac{\eps_{\gamma,\rm
pk}}{\eps_{\gamma}} \right)^2 \right] ;
\eps_{\gamma} < \frac{m_{l}^2 c^4 \Gamma^2}{2\eps_{\gamma,\rm pk}} 
\cr
\left( \frac{\eps_{\gamma,\rm pk}} {\eps_{l,\rm th}} -1 \right) +
\frac{1}{2} \left[ 1- \left( \frac{\eps_{\gamma,\rm pk}} 
{\eps_{\gamma}} \right)^2 \right] \cr ~~~~~~~;
\frac{m_{l}^2 c^4 \Gamma^2}{2\eps_{\gamma,\rm sa}} > \eps_{\gamma} > 
\frac{m_{l}^2 c^4 \Gamma^2}{2\eps_{\gamma,\rm pk}}
\cr
\frac{2}{3} \left( \frac{\eps_{\gamma,\rm pk}} 
{\eps_{\gamma,\rm sa}} \right)
\left( 1- \frac{\eps_{l,\rm th}} {\eps_{\gamma,\rm sa}} \right)^{3/2}
+ \left( \frac{\eps_{\gamma,\rm pk}} {\eps_{\gamma,\rm sa}} -1 \right)
\cr ~~~~~~~ + \frac{1}{2} \left[ 1- \left( \frac{\eps_{\gamma,\rm pk}} {
\eps_{\gamma}} \right)^2 \right] ;
\eps_{\gamma} > \frac{m_{l}^2 c^4 \Gamma^2}{2\eps_{\gamma,\rm sa}}.
\end{cases}
\label{opt-beta-1}
\ea
Here we defined the threshold energy for lepton pair production as
$\eps_{l,\rm th} = m_l^2 c^4 \Gamma^2/2\eps_{\gamma}$. Note that the
high energy photons produce $e^+e^-$ pairs dominantly at lower energy.
The ratio of the two opacities $\kappa = \tau_{\gamma\gamma
\ra \mu^+\mu^-} / \tau_{\gamma\gamma \ra e^+e^-}$ becomes unity for higher
energy photons, since the cross-section is the same for $\mu^+\mu^-$ and
$e^+e^-$ pair productions above the muon pair production threshold energy.

\subsection{Muon decay neutrino flux} 

Muon pairs decay to neutrinos as $\mu^- \ra e^- {\bar \nu_e}
\nu_{\mu}$ and $\mu^+ \ra e^+ \nu_e {\bar \nu}_{\mu}$ shortly after 
they are created in the c.m. frame of the $\gamma\gamma$ collision. In
the observer's frame $\eps_{\mu} \approx \eps_{\gamma}/2$, and the
particle pairs move radially along the incident photon's
direction. For simplicity we assume that the $\nu_e$ and $\nu_{\mu}$
created from $\mu$-decay carry 1/3 of the muon energy each.  The
observed neutrino energies are then $\eps_{\gamma}/6$ for each flavor.
The neutrino source flux, which is the same for $\nu_e$, $\nu_{\mu}$,
${\bar \nu}_e$ and ${\bar \nu}_{\mu}$ previous to any flavor
oscillation in vacuum, is
\ba 
\eps_{\nu}
\Phi_{\nu, \gamma\gamma}^{\rm s} \equiv \eps_{\nu}
\frac{dN_{\nu}}{d\eps_{\nu}} = {\rm min} [1, \tau_{\gamma\gamma \ra
\mu^+\mu^-}] \kappa \eps_{\gamma} \frac{dN_{\gamma}}{d\eps_{\gamma}}.
\label{gg-nuflux-define} 
\ea 

The high energy muons produced from $\gamma\gamma$ interactions may
lose a significant fraction of their energy by synchrotron radiation
before they decay into neutrinos (with a decay time $t_{\rm dec}$), if
their energy is above a break energy
\ba
\eps_{\mu,\rm sb} &=& 
\left( 6\pi m_{\mu}^5 c^5 \Gamma^2/
[t_{\rm dec} \sigma_{\rm Th} m_e^2 B'] \right)^{1/2}
\nonumber \\ &\approx & 5\times 10^7 \left(
\vareps_{e,-1.3} \Gamma_{2.5}^4 r_{14}^2/ 
[\vareps_{B,-1} L_{52}] \right)^{1/2} {\rm GeV}.~~~~~
\label{mu-break-energy}
\ea
The corresponding neutrino break energy from muon decay is
$\eps_{\nu,\rm sb} = \eps_{\mu,\rm sb}/6$. For $\eps_{\nu} \gtrsim
\eps_{\nu,\rm sb}$, the neutrino flux index would steepen by a factor
2 \cite{rachen98}.

\begin{figure} [ht]
\centerline{\epsfxsize=3.4in \epsfbox{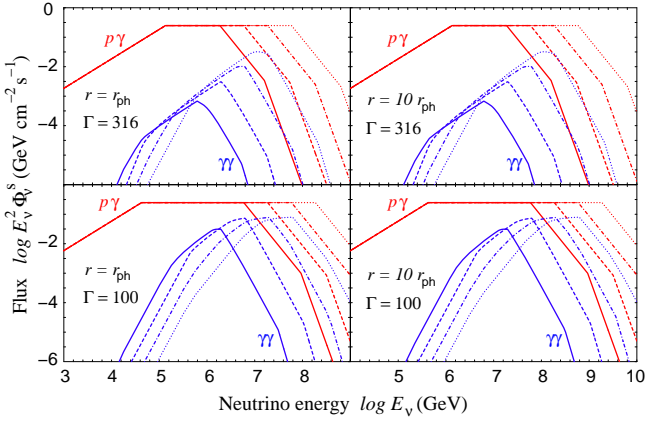}} \caption{ Source 
flux of $\nu_e$ (same for $\nu_{\mu}$, ${\bar \nu}_e$ and ${\bar
\nu}_{\mu}$) from $\gamma\gamma \ra \mu^+\mu^-$ interaction and
subsequent $\mu$-decays, compared to the canonical $\nu_e$ flux (same
for ${\bar \nu}_\mu$) from single $p\gamma \ra n\pi^+$ interactions
and subsequent $\pi^+$, $\mu^+$ decays. This is for a long duration
GRB of $L_{\gamma} = 10^{52}$ erg/s at redshift $z\sim 0.1$.  The
solid, dashed, dot-dashed and dotted lines are for magnetic field
parameters $\vareps_B = 10^{-1},~ 10^{-2},~ 10^{-3}$ and $10^{-4}$
respectively, for different $\Gamma$ and $r$ combinations.}
\label{fig:flux}
\end{figure}

We have plotted in Fig. \ref{fig:flux} the $\nu_e$ flux at the source,
$\eps_{\nu}^2 \Phi_{\nu}^{\rm s}$ (same for $\nu_e$, $\nu_{\mu}$,
${\bar \nu}_e$ and ${\bar \nu}_{\mu}$), previous to any vacuum
oscillation, arising from $\gamma\gamma \ra \mu^+ \mu^-$ interactions
and the associated muon decays, for a GRB of isotropic equivalent
luminosity $L_\gamma=10^{52}$ erg/s, which is average for a long GRB
\cite{zhang04}. We assume a redshift of $z\sim 0.1$, which is near the
low end of observed redshifts; there have been a few spectroscopic
redshifts observed in the $0.1-0.2$ in the past 8 years, and indirect
redshift measures, such as time lags, indicate many more bursts in
this range among the BATSE sample (see, e.g., Ref. \cite{zhang04} and
references therein).  Also plotted are the $\nu_e$ source flux:
$\eps_{\nu}^2 \Phi_{\nu}^{\rm s} = {\rm min} [1,\tau'_{p\gamma}] (0.2/8)
L_{\gamma}/(4\pi d_L^2 \vareps_e)$, from $p\gamma\ra \Delta^+ \ra
n\pi^+$ interactions and subsequent $\pi^+$ and $\mu^+$ decays.
Different panels are for different bulk Lorentz factor $\Gamma$ and
dissipation radii $r$.

\subsection{Neutrino flavor oscillation and flux on Earth} 

While neutrinos are created via weak interactions as flavor
eigenstates, their propagation is determined by the mass
eigenstates. The flavor eigenstates $\nu_{\alpha}$ and the mass
eigenstates $\nu_{j}$ are mixed through a unitary matrix defined as
$\nu_{\alpha} = \sum_{j} U^{*}_{\alpha j} \nu_{j}$, where
$\alpha=e,\mu,\tau$ and $j=1,2,3$ for three known flavors. The
probability for flavor change by vacuum oscillation is given by ${\cal
P}_{\nu_{\alpha} \ra \nu_{\beta}} = \sum_{j} \left|U_{\beta
j}\right|^2 \cdot \left|U_{\alpha j}\right|^2$, for the neutrino
propagation from their sources to Earth over astrophysical distances.

We use the standard expression for $U_{\alpha,j}$ with solar mixing
angle $\theta_{\odot} \equiv \theta_{12} = 32.5^{\circ}$ and
atmospheric mixing angle $\theta_{\rm atm} \equiv \theta_{23} =
45^{\circ}$ \cite{pdg04}. The unknown mixing angle $\theta_{13}$ and
the CP violating phase may be assumed to be zero given the current
upperbounds from reactor experiments. Using these values for
$U_{\alpha,j}$ and ${\cal P}_{\nu_{\alpha} \ra \nu_{\beta}}$ results
in a relationship between the source neutrino fluxes $\Phi_{\nu}^{\rm
s}$ and the expected neutrino fluxes on Earth $\Phi_{\nu}$ which is
given by
\ba
\left[ \begin{array}{c} \Phi_{\nu_e} \\ 
\Phi_{\nu_{\mu}} \\ \Phi_{\nu_{\tau}} \end{array} 
\right] \approx \left[ \begin{array}{ccc} 0.6 & 0.2 & 0.2\\ 
0.2 & 0.4 & 0.4\\ 0.2 & 0.4 & 0.4 \end{array} \right]
\left[ \begin{array}{c} \Phi_{\nu_e}^{\rm s} \\ 
\Phi_{\nu_{\mu}}^{\rm s} \\ \Phi_{\nu_{\tau}}^{\rm s}
\end{array} \right].
\label{nuflux-relation}
\ea
For antineutrinos ${\cal P}_{\bar \nu_{\alpha} \ra \bar \nu_{\beta}}$
is the same as above.

\begin{figure} [ht]
\centerline{\epsfxsize=3.4in \epsfbox{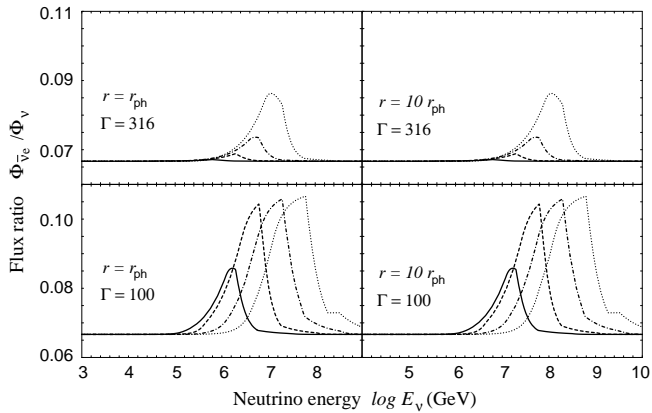}} \caption{ Observed 
ratio of electron antineutrino flux to the total neutrino flux
$\Phi_{\nu_e}/ \Phi_{\nu}$ for the $p\gamma$ and $\gamma\gamma$ source
fluxes plotted in Fig. \ref{fig:flux}. This is for a long duration GRB
of $L_{\gamma} = 10^{52}$ erg/s at redshift $z\sim 0.1$, the solid,
dashed, dot-dashed and dotted lines indicate magnetic field parameters
$\vareps_B = 10^{-1},~ 10^{-2},~ 10^{-3}$ and $10^{-4}$ respectively,
for different $\Gamma$ and $r$ combinations. Note that the 1/15 ratio
from single $p\gamma \ra n\pi^+$ interactions is enhanced by
$\gamma\gamma \ra \mu^+\mu^-$ interactions in certain energy ranges
which depend upon the GRB model parameters. }
\label{fig:fluxratio}
\end{figure}

Different production mechanisms produce $\nu$ and $\bar \nu$ fluxes at
the source with different flavor proportions. Their production ratios
may be expressed as normalized vectors, shown in the left hand side of
Eqs. (\ref{flux-ratios1} \& \ref{flux-ratios2}). The corresponding
flux ratios at Earth, using Eq. (\ref{nuflux-relation}), are shown in
the right hand side of Eqs. (\ref{flux-ratios1} \&
\ref{flux-ratios2}) below
\ba
\begin{array}{lccccc}
& \Phi_{\nu}^{\rm s} & \Phi_{\bar \nu}^{\rm s} & &
\Phi_{\nu} & \Phi_{\bar \nu} \cr p\gamma\ra n\pi^+
& \left[ \begin{array}{c} 1\\ 1\\ 0 \end{array} \right] & 
\left[ \begin{array}{c} 0\\ 1\\ 0 \end{array} \right] & 
\Rightarrow &
\left[ \begin{array}{c} 0.8\\ 0.6\\ 0.6 \end{array} \right] &
\left[ \begin{array}{c} 0.2\\ 0.4\\ 0.4 \end{array} \right],
\end{array}
\label{flux-ratios1}
\ea
\ba
\begin{array}{lccccc}
& \Phi_{\nu}^{\rm s} & \Phi_{\bar \nu}^{\rm s} & &
\Phi_{\nu} & \Phi_{\bar \nu} \cr \gamma\gamma\ra \mu^+\mu^-
& \left[ \begin{array}{c} 1\\ 1\\ 0 \end{array} \right] & 
\left[ \begin{array}{c} 1\\ 1\\ 0 \end{array} \right] & 
\Rightarrow &
\left[ \begin{array}{c} 0.8\\ 0.6\\ 0.6 \end{array} \right] &
\left[ \begin{array}{c} 0.8\\ 0.6\\ 0.6 \end{array} \right].
\end{array}
\label{flux-ratios2}
\ea
The source $\nu$-fluxes plotted in Fig. \ref{fig:flux} will be
modified accordingly. Note that $\Phi_{{\bar \nu}_e} (\gamma\gamma)/
\Phi_{{\bar \nu}_e} (p\gamma) \approx 4$ for the same initial 
$\gamma\gamma$ and $p\gamma$ flux levels. The ${\bar \nu}_e$-flux
component is 1/5 (1/15) of the total $\nu$-flux from $\gamma\gamma$
($p\gamma$). The observed ratio of ${\bar \nu}_e$ to total $\nu$ 
fluxes from both the single $p\gamma$ interactions and $\gamma\gamma$ 
interactions can be calculated using Eqs. (\ref{flux-ratios1} \&
\ref{flux-ratios2}) from the source fluxes as
\ba
\frac{\Phi_{{\bar \nu}_e}}{\Phi_{\nu}} = 
\frac{0.2 \Phi^{\rm s}_{{\bar \nu}_e, p\gamma} +
0.8 \Phi^{\rm s}_{{\bar \nu}_e, \gamma\gamma}}
{3 \Phi^{\rm s}_{{\bar \nu}_e, p\gamma} +
4 \Phi^{\rm s}_{{\bar \nu}_e, \gamma\gamma}}
\ea
We have plotted this ratio in Fig. \ref{fig:fluxratio} for different
GRB model parameters. The ratio is enhanced from the canonical
($p\gamma$) 1/15 value in the energy range where $\gamma\gamma$
interactions contribute significantly (see
Fig. \ref{fig:flux}). Interestingly, the enhancement takes place over
a small energy range which may be explored to learn about the GRB
model parameters as we discuss next.

\section{Neutrino detection} 

We consider here the anti-electron neutrino detection channel at the
Glashow resonance energy $\eps_{\nu, {\rm res}} = m_W^2 c^2/2m_e
\approx 6.4$ PeV \cite{glashow:60}. The number of electrons in the 2
km$^3$ effective IceCube volume is $N_{e,\rm eff} \approx 6\times
10^{38}$ and the corresponding number of downgoing ${\bar
\nu}_e$ events from a point source of flux $\Phi_{{\bar \nu}_e}$ is
\cite{Anchordoqui:2004eb}
\ba N_{{\bar \nu}_e} \approx \Delta t N_{e,\rm eff} \frac{\pi g^2
(\hbar c)^2}{4 m_e c^2} \Phi_{{\bar \nu}_e} ( \eps_{\nu, {\rm res}} ).
\label{resonant-events}
\ea
Here $g^2 \simeq 0.43$ from the standard model of electro-weak theory,
and $\Delta t$ is the duration of the emission.

\begin{figure} [ht]
\centerline{\epsfxsize=3.4in \epsfbox{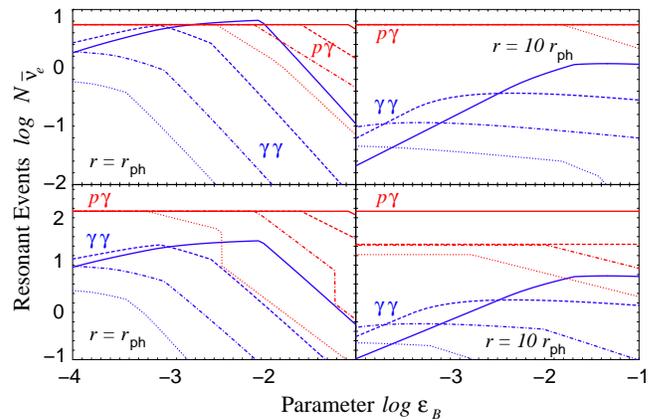}} \caption{
Resonant ${\bar \nu}_e$ events in IceCube from $\gamma\gamma$ and
$p\gamma$ interactions in a GRB, as a function of the magnetization
$\vareps_B$ for Lorentz factor $\Gamma$ of 100 (solid lines), 178
(dashed lines), 316 (dot-dashed lines), 500 (dotted lines) and
fireball radii equal to the photospheric radius ({\em two left
panels}) and ten times the photospheric radii ({\em two right
panels}). We used single $p\gamma$ interactions ({\em two top panels})
and multiple $p\gamma$ and subsequent $n\gamma$ interactions
proportional to $\tau_{p\gamma}$ ({\em two bottom panels}) for
comparison. Other GRB parameters are $z=0.1$, $\Delta t =30$ s,
$L_{\gamma} = 10^{52}$ erg/s and $\vareps_{e} = 0.05$.}
\label{fig:nue-event}
\end{figure}

We have plotted in Fig. \ref{fig:nue-event} the expected number of
${\bar \nu}_e$ events at $\eps_{\nu, {\rm res}}$ from a GRB fireball
for various value of $r$, $\vareps_B$ and $\Gamma$. We have assumed
here that shock accelerated protons interact once with synchrotron
photons (two top panels), losing $\sim 20 \%$ of their energy. For
very high $p\gamma$ opacity the protons can lose most of their energy
through the $p\gamma$ and $n\gamma$ interaction chains. This could
lead to the $\nu_e$-fluxes from both $p\gamma$ and $\gamma\gamma$
plotted in Fig. \ref{fig:flux} to roughly increase by a factor
five. The $p\gamma$ source $\nu$-fluxes may reach ratios
$\Phi_{\nu}^{\rm s} = \Phi_{\bar \nu}^{\rm s} = [1,2,0]$, from $\pi^+$
and $\pi^-$ decays, which at Earth would be $\Phi_{\nu} = \Phi_{\bar
\nu} = [1,1,1]$ for $\eps_{\nu} \lesssim \eps_{\mu,\rm sb}/2$. The
resonant ${\bar \nu}_e$-events in such case, from both $\gamma\gamma$
and $p\gamma$ fluxes, are plotted in Fig. \ref{fig:nue-event} (two
bottom panels).  For simplicity, we assumed $\Phi_{\nu} \propto
\Phi_{\gamma} \propto {\rm min}[1,0.2 \tau'_{p\gamma}]$ for the $p\gamma$ 
$\nu$-flux and $\gamma$-flux in Eq.  (\ref{gg-nuflux-define}).  Also
$\Phi_{\bar \nu} = [1,1,1]$ from $p/n \gamma$ interactions for
$\tau'_{p\gamma} \ge 2$.

The background for astrophysical ${\bar \nu}_e$ detection is mostly
due to atmospheric prompt neutrinos from cosmic-ray generated charm
meson decays. A parametrization for the $\nu_e$ and ${\bar
\nu}_e$ atmospheric flux is given by \cite{thunman96} 
\ba
\Phi_{\nu_e +{\bar \nu}_e}^{\rm atm} &=& \begin{cases} 
\frac{1.5\times 10^{-5} \eps_{\nu}^{-2.77}} 
{1+ 3\times 10^{-8} \eps_{\nu}}; \eps_{\nu} < 1.2\times 10^{6}~{\rm
GeV} \cr \frac{4.9\times 10^{-4} \eps_{\nu,{\rm ob}}^{-3.02}} {1+
3\times 10^{-8} \eps_{\nu}}; \eps_{\nu} > 1.2\times 10^{6}~{\rm GeV}
\end{cases} \nonumber \\ && ~\times 
~{\rm GeV}^{-1} ~{\rm cm}^{-2} ~{\rm s}^{-1} ~{\rm sr}^{-1}.
\label{atmoflux}
\ea
The corresponding background ${\bar \nu}_e$-events at the Glashow
resonance energy is $\lesssim 10^{-7}$ for a GRB within a $\sim 100$ s
time window, allowing the full directional uncertainty ($2\pi$ sr),
given the poor current knowledge of the $\nu_e$ or ${\bar \nu}_e$
signal reconstruction in neutrino Cherenkov detectors, from
Eq. (\ref{resonant-events}).

\section{Discussion} 

The results of Fig. \ref{fig:nue-event} and Eqs. (\ref{flux-ratios1}
\& \ref{flux-ratios2}) show that $\gamma\gamma$ interactions in
astrophysical sources can enhance the observed
$\Phi_{\bar{\nu}_e}:\Phi_\nu$ flux ratio. A different source of
enhancement of the ${\bar \nu}_e$ flux may be $pp$ interactions
\cite{Anchordoqui:2004eb,kashti05}. In GRBs, however, the optical
depth to $pp$ is low \cite{wb97}, except in buried jets leading to
$\nu$ precursors \cite{razzaque03}, where $pp$ interactions are
expected to lead to $\nu$'s at energies $\sim$ TeV. The number of
resonant ${\bar \nu}_e$ events arising from $p\gamma$ interactions is
essentially independent of $\vareps_{B}$ (for $\vareps_{B}\lesssim
10^{-2}$) for any $\Gamma$. On the other hand, the number of resonant
${\bar \nu}_e$ events arising from $\gamma\gamma$ interactions varies
significantly with $\Gamma$ and $r$. It may become as large as the
$p\gamma$ contribution for $10^{-2} \lesssim \vareps_{B} \lesssim
10^{-3}$, $100 \lesssim \Gamma \lesssim 300$ and $r_{\rm ph} \lesssim
r \lesssim 3 r_{\rm ph}$.  For long bursts of average
isotropic-equivalent luminosity at a redshift $\sim 0.1$, which from
past experience are electromagnetically detected every few years,
IceCube could probe the ${\bar \nu_e}$ enhancement, and thus the value
of the magnetization parameter and dissipation radius, by measuring
the $\Phi_{\bar{\nu}_e}:\Phi_\nu$ flux ratio. Finally, we note that a
moderate excess of $\nu_e$ events compared to $\nu_{\mu}$ and
$\nu_{\tau}$ events may also be an indication for the presence of a
$\gamma\gamma$ component.

\section*{Acknowledgements} 

Work supported by NSF grant AST0307376. EW is partly supported by
Minerva and ISF grants.

\end{document}